# Authentication Modeling with Five Generic Processes

Sabah Al-Fedaghi[1], MennatAllah Bayoumi[2]
Computer Engineering Department
Kuwait University, Kuwait

*Abstract*—Conceptual modeling is an essential tool in many fields of study, including security specification in information technology systems. As a model, it restricts access to resources and identifies possible threats to the system. We claim that current modeling languages (e.g., Unified Modeling Language, Business Process Model and Notation) lack the notion of *genericity*, which refers to a limited set of elementary processes. This paper proposes five generic processes for modeling the structural behavior of a system: creating, releasing, transferring, receiving, and processing. The paper demonstrates these processes within the context of public key infrastructure, biometric, and multifactor authentication. The results indicate that the proposed generic processes are sufficient to represent these authentication schemes.

## I. INTRODUCTION

Security is a necessary feature in information technology (IT) systems. Security specification requires identifying risks, access requirements, and recovery strategies, and comprises well-developed security mechanism processes [1]. Early-stage development of security specification assists in lowering the possibility of security breaches.

Authorization and authentication both play vital roles in the configuration of security mechanisms. Authorization is the process of allowing users to access system objects based on their identities. Authentication confirms that the user is who he or she claims to be.

Conceptual modeling is a description of reality using a modeling language to create a more-or-less formalized schema [2]. A conceptual model in the security field restricts access to the resources and identifies possible threats to the system. In modeling, *notations* (diagrams, symbols, or abbreviated expressions) are required to specify technical facts and related concepts of systems. They are necessary to articulate complex ideas succinctly and precisely [3]. For a notation to convey accurate communication, it must effectively represent the different aspects of a system and be well understood among project participants. The historic roots of modeling notations in software engineering can be traced back to structured analysis and design, which are based on data flow diagrams [3].

### A. Security Modeling

Many languages and mechanisms, such as Business Process Model and Notation (BPMN) [4], secure Tropos [5], misuse cases [6], and mal-activity diagrams [7], are used in the field of security modeling. For space consideration, we focus here on the Unified Modeling Language (UML) and BPMN.

The UML [8] has been utilized as a graphical notation to construct and visualize security aspects in object-oriented systems. It is currently utilized as a primary notation for security and authentication because it provides a spectrum of notations representing the various aspects of a system. The use of the UML for conceptual modeling requires special care to not confuse software features with aspects of the real world being modeled [9].

BPMN was designed to be used by people without much training in software development. "UML diagrams look technical, and in practice, they are much harder for businesspeople to understand than BPMN diagrams" [10]. BPMN includes a rich set of model constructs for business process modeling.

This paper is about conceptual modeling. It is part of a research project that applies a new modeling language, the thing machine (TM), to modeling computer attacks [11]. The paper concentrates on using the TM to model authentication. The thesis promoted in our research works is that modeling in the abovementioned languages lacks *genericity*, a notion for representing systems that forms the base for process modeling. This has caused conceptual vagueness that obstructs the differentiation of objects. A specific goal of the paper is to substantiate the viability of the TM by applying it to modeling authentication.

### B. Modeling Authentication

In the twenty-first century, few matters are more pressing than those related to identity authentication. Authentication is a mechanism used to make sure that those obtaining session access are who they say they are. To access online systems and services, we all face the challenge of proving our identities [12].

In the real world, thousands have found themselves blocked from opening bank accounts, making payments, or travelling because of an unfortunate name similarity to those individuals or entities on a sanctions list. Hundreds of thousands have been victims of identity fraud, often only learning of the crime when they apply for credit and find their credit rating has been compromised by fraudulent loans obtained in their names [12].

In this paper, we focus on individual and entity authentication for digital interactions. We concentrate on authentication in the context of usability of IT systems in terms of who is using the system, what they are using it for, and the environment in which they are using it (ISO standard 9241 Part 11). The ISO 9241 standard for identity authentication is made up of three components: what you are (e.g., biometric information), what you have (e.g., having a token), and what you know (e.g., PINs, passwords).

### C. Examples of Modeling Authentication

Fig. 1 shows a typical authentication process—in this case, a single sign-on (SSO) that allows a user to access multiple applications with one set of login credentials. This SSO is modeled using a UML activity diagram. With an SSO, a client accesses multiple resources connected to a local area network [13]. Partner companies act as identity providers and control usernames and other information used to identify and authenticate users for web applications. Each partner provides Google with the URL of its SSO service, as well as the public key that Google will use to verify Security Assertion Markup Language (SAML, a protocol that refers to what is transmitted regarding identity information between parties) responses. When a user attempts to use a hosted Google application, Google generates an SAML authentication request and sends a redirect request back to the user's browser that points to the specific identity provider. The SAML authentication request contains the encoded URL of the Google application that the user is trying to reach. This authentication process continues as shown in part in Fig. 1 [13]. Nevertheless, in general, "activity diagrams have always been poorly integrated, lacked expressiveness, and did not have an adequate semantics in UML" [14]. With further development of the UML, "several new concepts and notations have been introduced, e.g., exceptions, collection values, streams, loops, and so on" [14].

Plavsic and Secerov [15] model the classic login procedure using different kinds of UML diagrams: deployment diagrams, use case diagrams, interaction overview diagrams, sequence diagrams (see Fig. 2), and class diagrams. The code shown in Fig. 2 was generated from the class diagrams. According to Plavsic and Secerov [15], the use case diagram counts as a starting point; however, use cases do not describe system structure or details of behavior. Therefore, Plavsic and Secerov use other diagrams to follow messages that are exchanged between objects and realize system functionality.

Lee [3] gives an example wherein field officers are required to provide authentication before they can use a system called FRIEND. Authentication is modeled as an authenticate-use case. Later, two more use cases are introduced: AuthenticateWithPassword, which enables field officers to login without any specific hardware, and AuthenticateWithCard, which enables field officers to log in using smart cards. The two use cases are represented as specializations of the Authenticate use case (see Fig. 3).

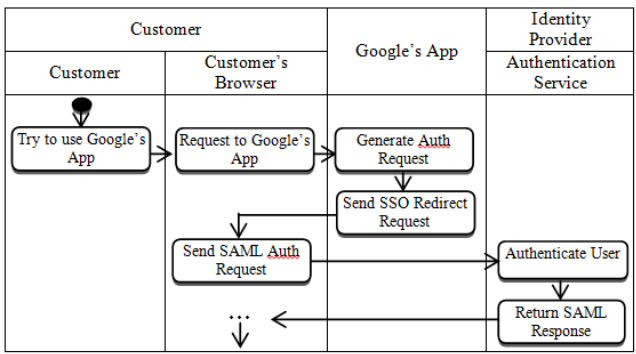

Fig. 1. An Example of a UML Activity Diagram for SSO to Google Apps. (Partially Redrawn from [13]).

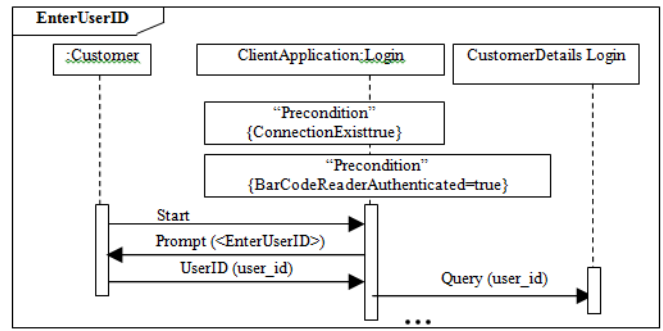

Fig. 2. EnteruserId Sequence Diagram. (Partially Redrawn from [15]).

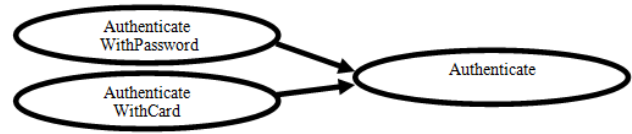

Fig. 3. The Authenticate use Case is a High-Level use Case Describing, in General Terms, the Process of Authentication. AuthenticatewithPassword and AuthenticatewithCard are Two Specializations of Authenticate. (Partially Redrawn from [3]).

In the next section, we will briefly review the TM with a new contribution related to the notion of genericity. In Section 3, we give an example. In Section 4, we apply the TM to model authentication.

## II. THING MACHINE WITH FIVE GENERIC PROCESSES

We claim that a modeling methodology is based on five generic (a notion to be discussed later) processes: creating, releasing, transferring, receiving, and processing (changing). These elementary processes form a complex abstract machine called a Thing Machine, as shown in Fig. 4. Fig. 5 shows a TM formulated to align with the classical input–process–output model.

The machines constitute a mosaic or network of machines. Additionally, the TM model embraces *memory* and *triggering* (represented as dashed arrows), relations among the processes' stages (machines). A TM manifests structure and behavior simultaneously. Only five elementary processes are used because they represent genericity in operation, the way the three states of water (liquid, vapor, and solid) represent three generic concepts. These elementary processes have been called different names.

- Create: generate, produce, manufacture, give birth, initiate, assemble, emerge, appear (in a system), etc. Process: change, modify, adjust, amend, etc.

- Receive: obtain, accept, collect, take, get, etc.

- Release: allow, relieve, discharge, let, free, etc.

- Transfer: transport, transmit, carry, communicate, etc.

The TM model has been applied to many real systems, such as phone communication [16], physical security [17], vehicle tracking [18], unmanned aerial vehicles [19], and programming [20].

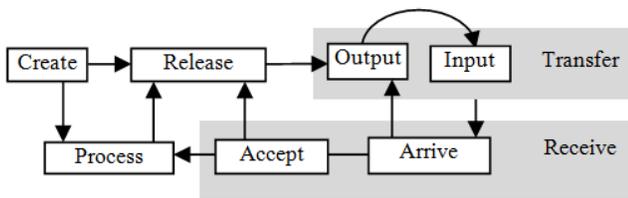

Fig. 4. Thinging Machine.

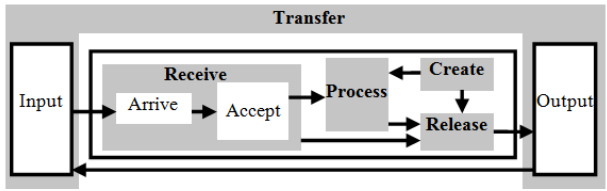

Fig. 5. Another form of Description of a TM.

### A. Philosophical Foundation of Things and Thinging

The TM model is constructed on the philosophical foundations of Heidegger's notions of a *thing* and *thinging* [21]. Heidegger's philosophy gives an alternative analysis of "(1) eliciting knowledge of routine activities, (2) capturing knowledge from domain experts and (3) representing organizational reality in authentic ways" [22]. Additional information about the philosophical foundations of the TM can be found in Al-Fedaghi [23-24].

Briefly, in a TM, a thing is defined as that which can be created, processed, released, transferred, and received. It encounters humans through its givenness (Heidegger's term). "In contrast to object-orientation, which represents things as quantifiable objects to be controlled and dominated, Heidegger's definition of a thing encompasses a particular concrete existence along with its interconnectedness to the world" [25]. According to Heidegger [21], thinging expresses how a "thing things," which he explained as gathering or tying together its constituent parts.

A TM operates by creating, processing, receiving, releasing, and transferring things. For example, a *tree* is a machine "through which flows of sunlight, water, carbon dioxide, minerals in the soil, etc. flow. Through a series of operations, the machine transforms those flows of matter, those other machines that pass through it into various sorts of cells" [26]. In the TM approach, a thing is not just an entity; it is also a machine that handles other things.

### B. Genericity

The TM's five processes are categorical. Members of each category have the following features:

- They focus on essential properties and ignore variations in the created category—for example, newness regardless of who, what, how, etc.

- They capture the blueprint aspect: e.g., creation is a "popping up" phenomenon wherein a thing either "emerges into" the system or as a result of existing things being processed to trigger the creation of other things.

- Things have attributes similar to *objects*—a created thing comes to "life", and a processed (changed) thing remains the same thing.

- Things have actions similar to *subjects* (machines)—creating a thing is "bringing it to life" and processing a thing changes it in some way.

A sketch of the proof of the necessity for the five generic processes can be outlined as shown in Fig. 6. Thus, informal justification for the five TM stages can be specified as follows:

- Things become entities in the system either by being imported from the outside (transfer/input) or by being internally constructed (creation). See Fig. 6(a).

- Things coming from the outside are either rejected from or received (receive) into the system. See Fig. 6(b).

- Things may flow outside the system (transfer/output). See Fig. 6(c).

- Deported things may be queued before transfer (release). See Fig. 6(d).

- Things inside the system may be processed (process). See Fig. 6(e).

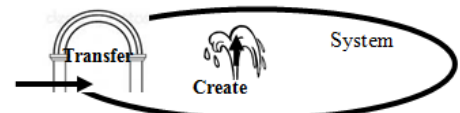

(a) Things become Entities in the System Either by being Imported from the Outside or by being Constructed Internally.

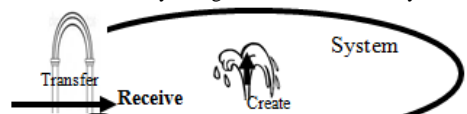

(b) Things Coming from the Outside are Either Rejected from or Received into the System.

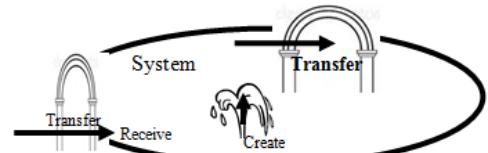

(c) Things may flow Outside the System.

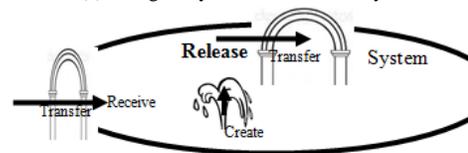

(d) Deported things may be Queued before Transfer.

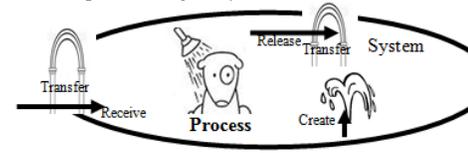

(e) Things Inside the System may be Processed.

Fig. 6. Informal Sketch of the Generic Processes.

This is what we mean by *generic* processes. Even though they are used differently according to the setting, members of each generic process seem to be synonymous with respect to *things*. In language, such a phenomenon appears in the case of the adjectives *big*, *great*, and *large*, which are seemingly synonymous words but are likely to be used in different ways in different settings [27]. Processes recognized as being of the same kind of "meaning" in the above sense are said to possess a generic property. Generic processes are conduits through which various types of processes flow.

### III. THING MACHINE MODELING EXAMPLE

Guizzardi and Wagner [2] give an example of a service queue system in which customers arrive at random times at a service desk. They have to wait in a queue when the service desk is busy. Otherwise, when the service desk is not busy, they are immediately served by the clerk. Whenever a service is completed, the next customer from the queue (if any) is served [28].

Fig. 7 shows the TM model of the example. The customer arrives (circle 1) to get into the queue (Q). We assume a circular queue structure stored in Q(0:n - 1) with mod n operation; *rear* points to the last item and *front* is one position counterclockwise from the first item in Q. As typically described, the queue has a *rear*, which, upon the arrival of the customer (2), is retrieved/released (3) and incremented (4). Hence:

- If Q is full (the maximum capacity of the queue when *(rear+1)mod n =front*), the system blocks any newly arriving customers.

- The new rear value is stored (6).

Accordingly, the customer is assigned a position (given a number) in the queue and joins the other customers waiting in the queue (8).

Whenever the service agent is *not busy* (9):

- The first customer in the queue is released to the service area (10).

- The arrival of the customer to the service area changes its state to *busy* (11).

- The customer is then processed (12).

- The customer is released (13), which triggers the not busy state (14).

- The customer leaves the service area (15).

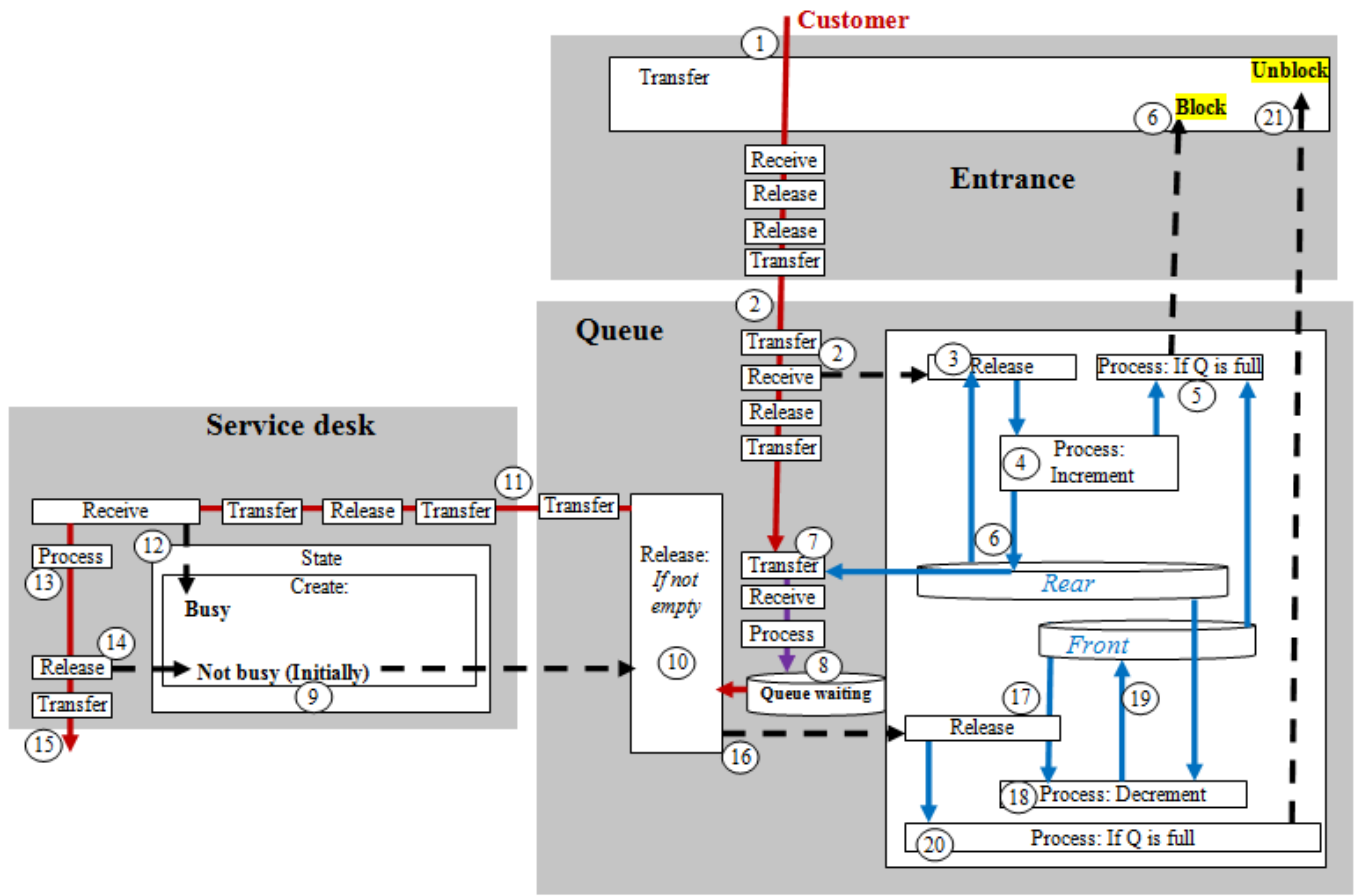

Fig. 7. Static TM Description of the Example.

Triggering the *not busy* state results in taking a new customer from the queue, as mentioned previously (10), and also updates the queue data (16). Thus,

- The front value is retrieved (17) and decremented (18), and the new value is stored (19).

- The original *front* value (before decrementing it) is checked (20), and if the Q was full, the blockage of new customers from entering the queue is lifted (21).

Initially, we assume that the entrance is not blocked, the queue is empty, and the service is not busy.

The dynamic behavior of the system can be developed based on events. An event in a TM is treated as a thing/machine—that is, it can be created, processed, released, transferred, and received. For example, the event a customer moves from the queue to the service desk is represented as shown in Fig. 8. It has two submachines: time and region where the event takes place. An event also denotes a change. All stages in the static description of Fig. 7 indicate elementary changes; however, we are typically interested in larger events that include several stages, as demonstrated in the event a customer moves from the queue to the service. Accordingly, we identify the following events in this example (see Fig. 9):

Event 1 ($E_1$): The service is open.
Event 2 ($E_2$): The service is closed (blocked).

Event 3 ($E_3$): A customer joins the queue.
Event 4 ($E_4$): Top is retrieved and incremented, and the new value is stored.
Event 5 ($E_5$): The queue is full (i.e., new value = max).
Event 6 ($E_6$): The queue is not full.
Event 7 ($E_7$): A customer joins the queue.
Event 8 ($E_8$): The service agent is not busy.
Event 9 ($E_9$): A customer moves from the queue to the service.
Event 10 ($E_{10}$): The service becomes busy.
Event 11 ($E_{11}$): The customer leaves the service.
Event 12 ($E_{12}$): Top is retrieved and decremented, and the new value is stored.
Event 13 ($E_{13}$): Top becomes less than max.

Fig. 10 shows the behavior of the system in terms of the chronology of its events.

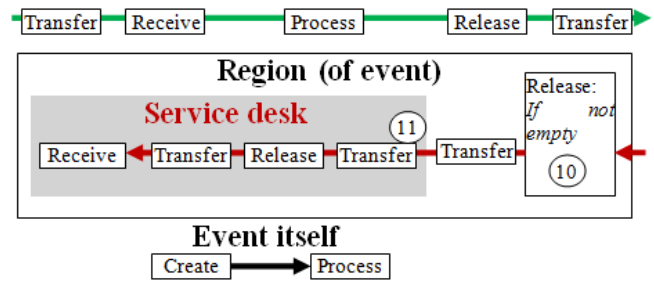

Fig. 8. Event with Region and Time Submachines.

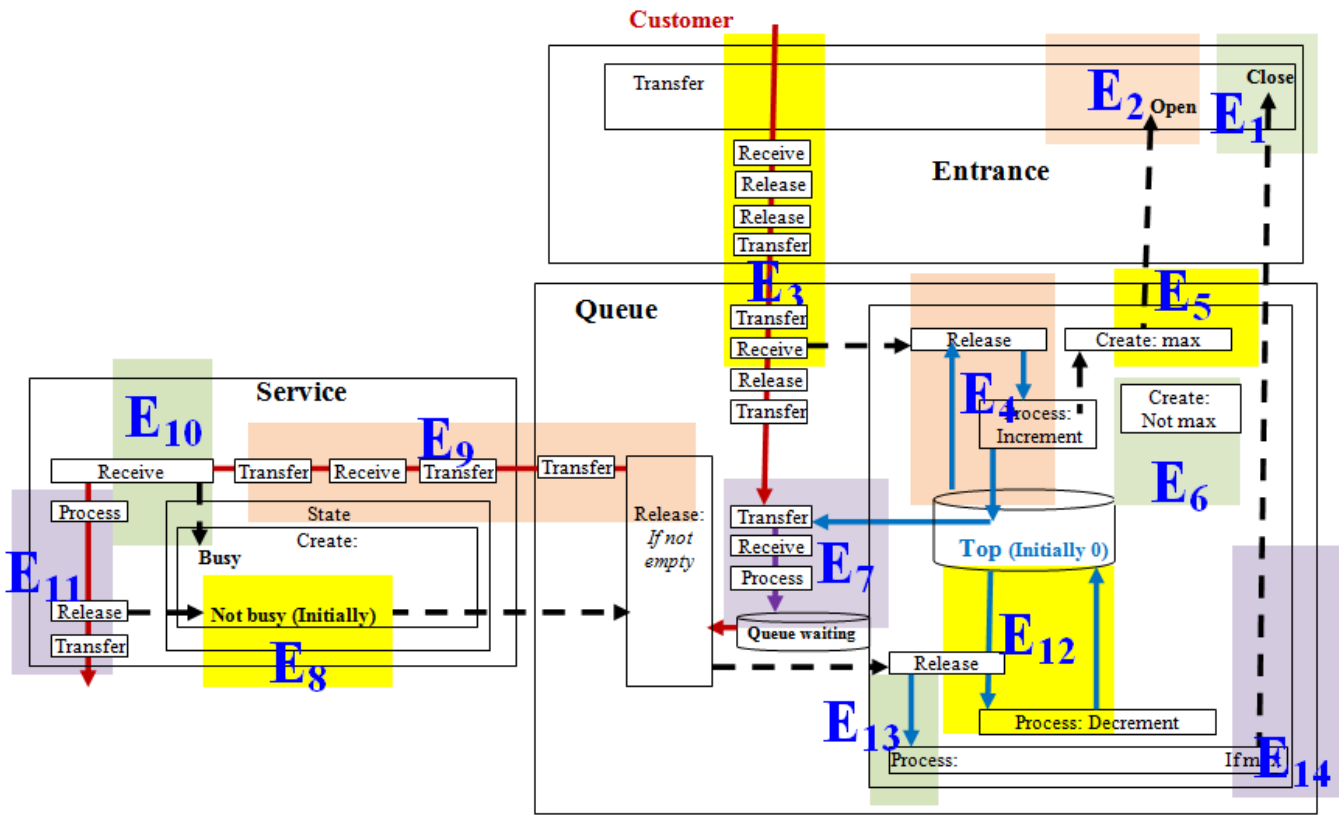

Fig. 9. Identifying the Events in the Static Description of the Example.

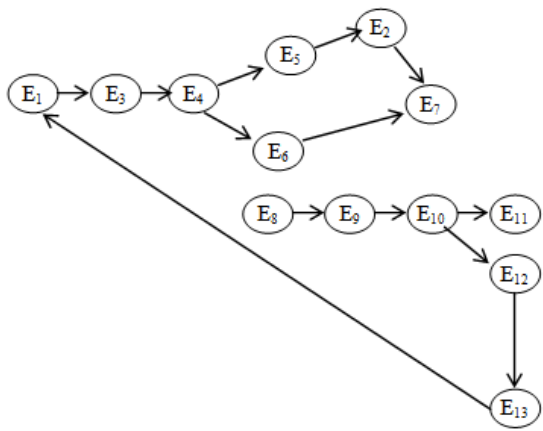

Fig. 10. Chronology of Events.

## IV. CASE STUDY: MODELING AUTHENTICATION

To apply a TM to modeling authentication, we adopt a security case study that involves insider attackers as presented by Nostro et al. [29]. This case study is interesting because it adopts a modeling approach using UML diagrammatic and textual use cases in line with the level of modeling applied in this paper. Additionally, UML use cases give us an opportunity to contrast use case diagrams with TM diagrams.

The case study includes the taxonomy of users physically or logically involved within the system and investigates their roles as potential insiders. The users are system administrator (SA), system expert, unknown user, domain expert, human sensor, and operator. Nostro et al. [29] explore only the SA and system expert, and we, in this paper, focus on the SA performing a *software update*. Fig. 11 shows the use case related to the SA; the darkened part indicates our region of emphasis. Fig. 12 shows the textual description of the use case.

Based on such a use case model that "guides the whole process," Nostro et al. [29] identify and assess insider threats and develop countermeasures that are oriented toward prevention, deterrence, or detection. They also use an ad hoc attack execution graph called ADVISE (see Fig. 13).

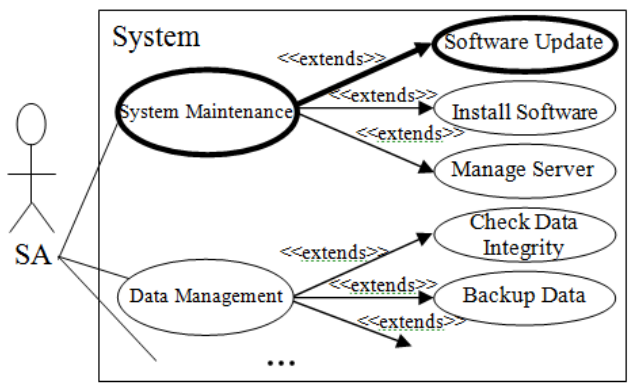

Fig. 11. UML use Case Diagram Involving the SA. (Partially redrawn from [29]).

| System Maintenance Use Case |
|---|
| *Actor/s*: SA |
| *Pre-condition*: The SA must be **authenticated.** |
| *Post-condition*: The SA has full access to the system. |
| *Description*: |
| Apply OS patches and upgrades on a regular basis to the system, the administrative tools, and utilities. Configure/add new services as necessary. Upgrade and configure system software or asset management applications. Maintain operation, configuration, or other procedures |
| ... |
| **Data Management** |
| *Actor/s*: SA ... |

Fig. 12. Description of UML use Case Diagram—SA. (Partially Taken from [29]).

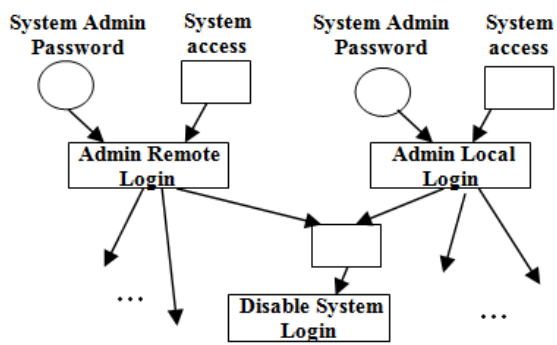

Fig. 13. Sample Attack Execution Graph. (Partially Redrawn from [29]).

We claim in this paper that the TM model presents a systematic alternative (one kind of notion) in modeling security. Without loss of generality, we will focus on the authentication part of Nostro et al. [29] to demonstrate the viability of the TM model.

## V. MODELING AUTHENTICATION

Authentication plays an important role in the security of computing, hence the existence of several authentication techniques. An authentication process attempts to verify a user's identity prior to the user's access to any resources in order to protect the system against various attack types. Once authenticated, the user is permitted to connect with cloud servers to request services [30-33]. Without loss of generality and due to space limitations, we will apply the TM model to only three authentication methods: public key infrastructure (PKI) authentication, biometric authentication, and multifactor authentication. As discussed in the case study in Section IV, the authentication of the SA is a precondition of all four use cases (system maintenance, data management, profile management, and crisis management, as represented in Fig. 11). The login session allows the SA to begin requesting services from the system. However, no requests from any of these four use cases will be serviced until the SA is authenticated by the system.

The first SA role to be investigated is the system maintenance case. This case is an umbrella to three subcases involving software updates, installing software, and managing servers.

### A. Public Key Infrastructure Authentication

Fig. 14 shows the TM representation of SA roles under the PKI framework system, whereas Fig. 15 shows the corresponding dynamic system, assuming the SA is already certified. The figure comprises two main machines: the SA and the system (highlighted in yellow).

- The SA logs into his or her account (Circle 1 in Fig. 14).

- Assuming correct credentials, the system creates (2) a session.

- The SA issues a request (3) for system maintenance, such as a software update.

- Upon receiving the request, the system performs the authentication process (4) [34] as follows:

  o The system generates random data (5) using the SA's public key and sends it to him or her (6).

  o The SA processes (7) the random data using his or her private key (8) and sends its encrypted version to the system (9).

  o The system uses the SA's public key (10) to decrypt (11) the incoming encrypted data, producing decrypted data (12).

  o The decrypted data are compared (13) to the original random data; if they are equivalent, a system maintenance session is opened for the SA (14).

A selected set of events are described as follows (see Fig. 15):

Event 1 ($E_1$): The SA logs into his or her account, and the system creates a session accordingly.
Event 2 ($E_2$): The SA issues a request to maintain the system.
Event 3 ($E_3$): The system starts the authentication process by generating random data and sending it to the SA.
Event 4 ($E_4$): The SA processes the random data using his or her private key and sends the encrypted data to the system.
Event 5 ($E_5$): The system uses the SA's public key to decrypt the incoming encrypted data, producing a decrypted dataset.
Event 6 ($E_6$): The original random data are transferred to the comparison module.
Event 7 ($E_7$): The decrypted data are compared to the original random data.
Event 8 ($E_8$): If the data are equivalent, a system maintenance session is opened for the SA.

Fig. 16 shows the chronology of these events that model the behavior of the PKI-based authentication system.

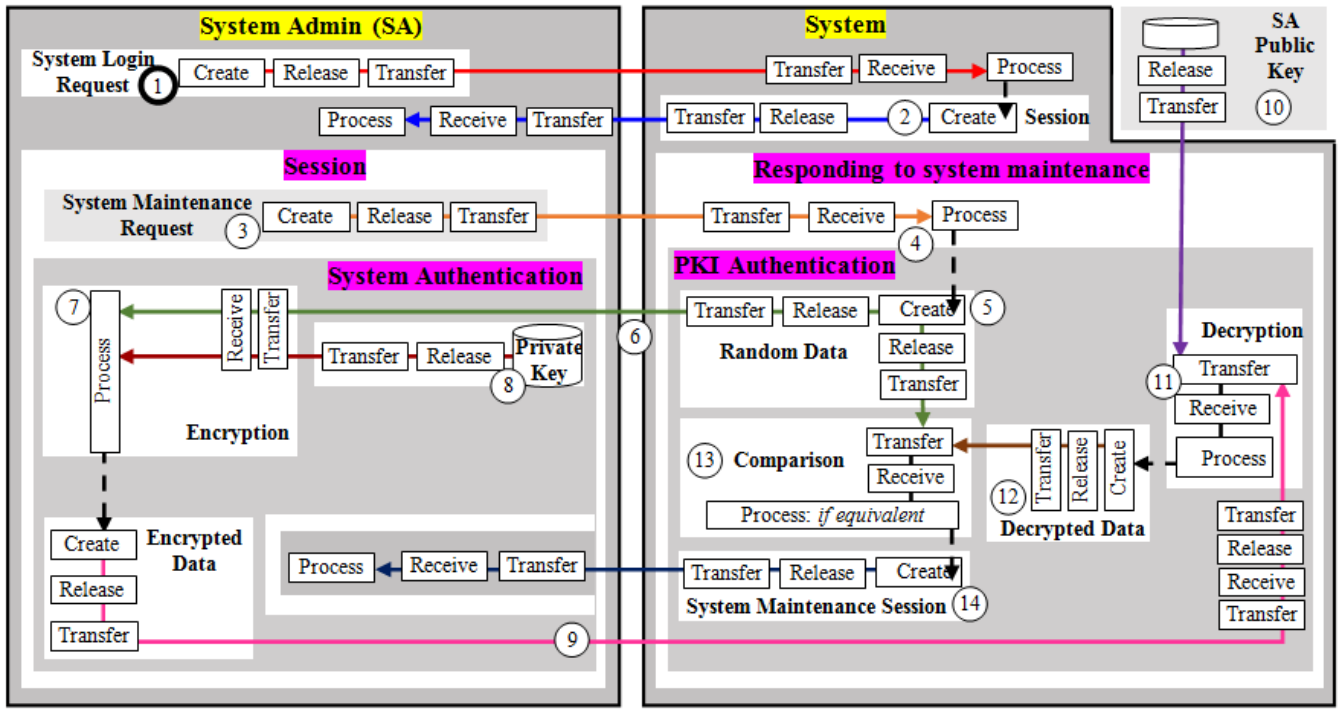

Fig. 14. TM Representation of UML use Case Involving the SA in PKI Authentication.

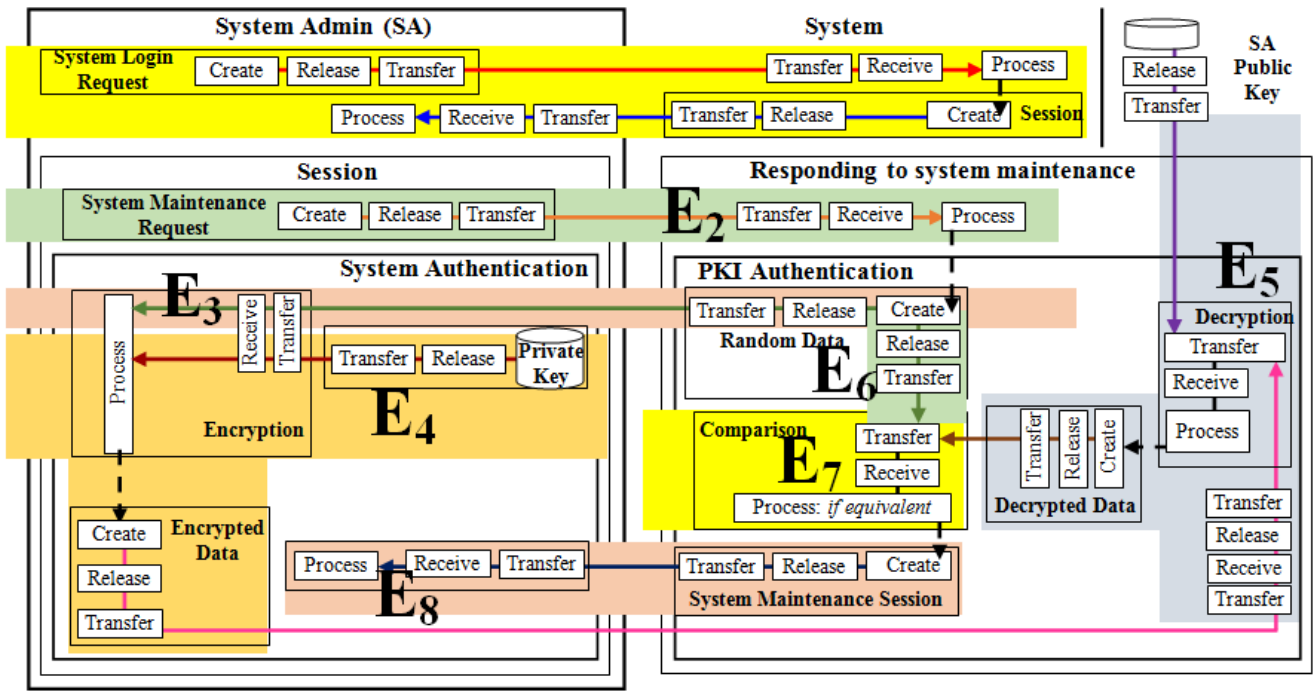

Fig. 15. Meaningful Events During PKI Authentication.

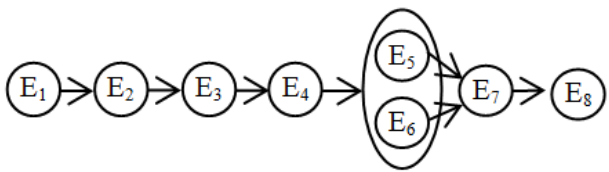

Fig. 16. Control of the PKI Event Sequence.

### B. Biometric Authentication

Fig. 17 and 18 show the static and dynamic TM representations of the SA's roles under a physical biometric authentication system. A typical physical biometric system carries out authentication in two stages—the enrollment stage and the verification stage.

Fig. 17 comprises two main machines: the SA and the system (highlighted in yellow).

- Initially, in the enrollment stage, the SA requests (1) the biometric trait desired, such as a face or fingerprint.

- In response, the system requests (2) the SA to present his or her chosen biometric trait.

- The SA then presents (3) the trait to the scanning hardware.

- The system then extracts (4) the scanned trait for encryption and storage (5).

- To initiate an interaction with the system, the SA logs into his or her account (6). With the correct credentials, the system creates (7) a session.

- The SA issues a request (8) to maintain the system (e.g., software update).

- The system starts the authentication process (9) (verification stage) [35].

  o The system requests (10) the SA to present his or her chosen biometric trait.

  o The SA then presents (11) the trait to the scanning hardware.

  o The system then extracts (12) the scanned trait for comparison purposes.

  o The originally encrypted trait is decrypted (13) and compared with the trait extracted from the scanning hardware (14). If they are equivalent, a system maintenance session is opened to the SA (15).

Fig. 18 shows the dynamic description of the model. A selected set of events is described as follows:

Event 1 ($E_1$): The SA requests the biometric trait desired for the enrollment stage, and the system requests the SA to present the chosen biometric trait.

Event 2 ($E_2$): The SA presents the trait to the scanning hardware for extraction.

Event 3 ($E_3$): The extracted data are then encrypted and stored.

Event 4 ($E_4$): The SA logs into his or her account, and the system creates a session accordingly.

Event 5 ($E_5$): The SA issues a request for system maintenance.

Event 6 ($E_6$): The system starts the authentication process by requesting the SA to present the chosen biometric trait.

Event 7 ($E_7$): The SA presents the trait to the scanning hardware for extraction.

Event 8 ($E_8$): The system decrypts the originally encrypted trait.

Event 9 ($E_9$): The extracted trait is compared to the decrypted data.

Event 10 ($E_{10}$): If the data are equivalent, a system maintenance session is opened to the SA.

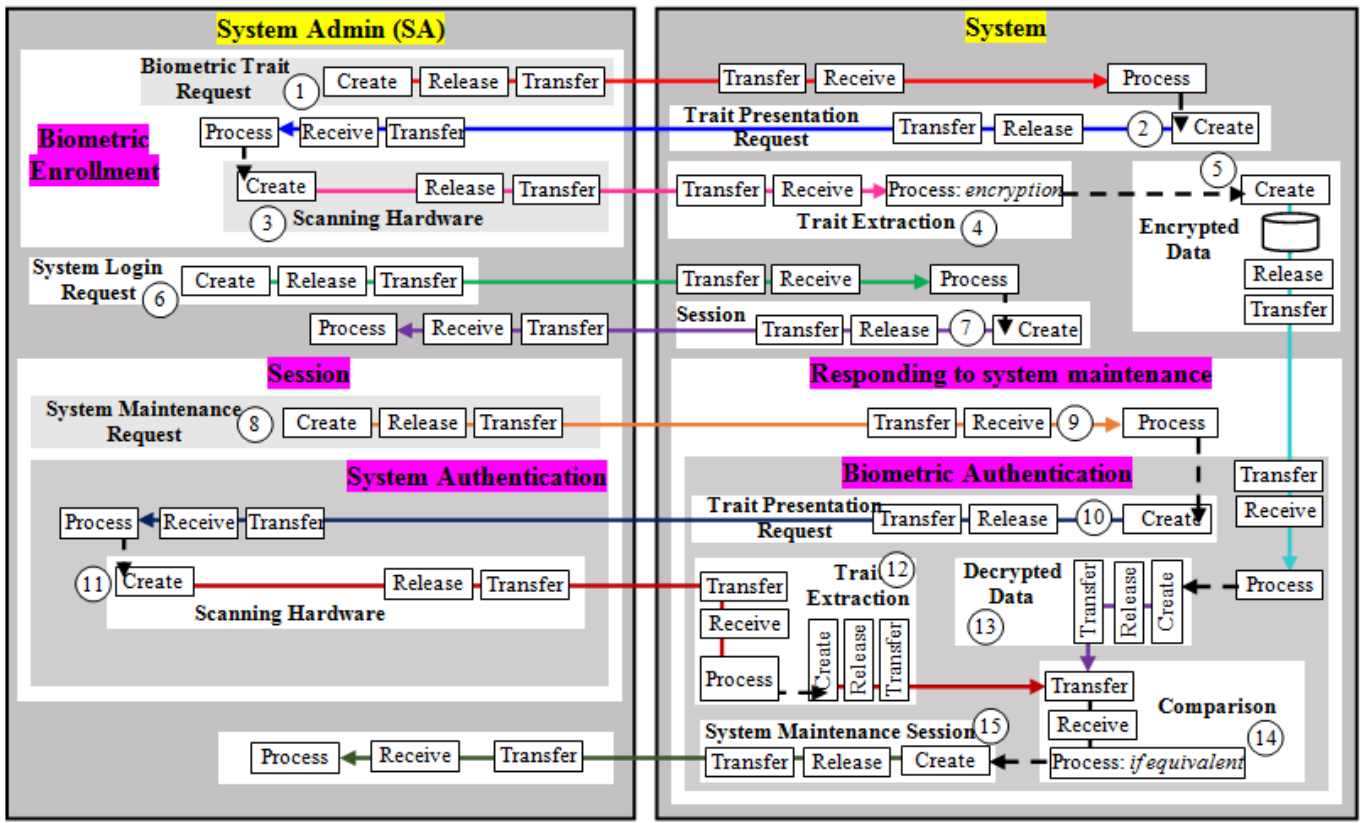

Fig. 17. TM Representation of UML use Case Involving the SA in Physical Biometric Authentication.

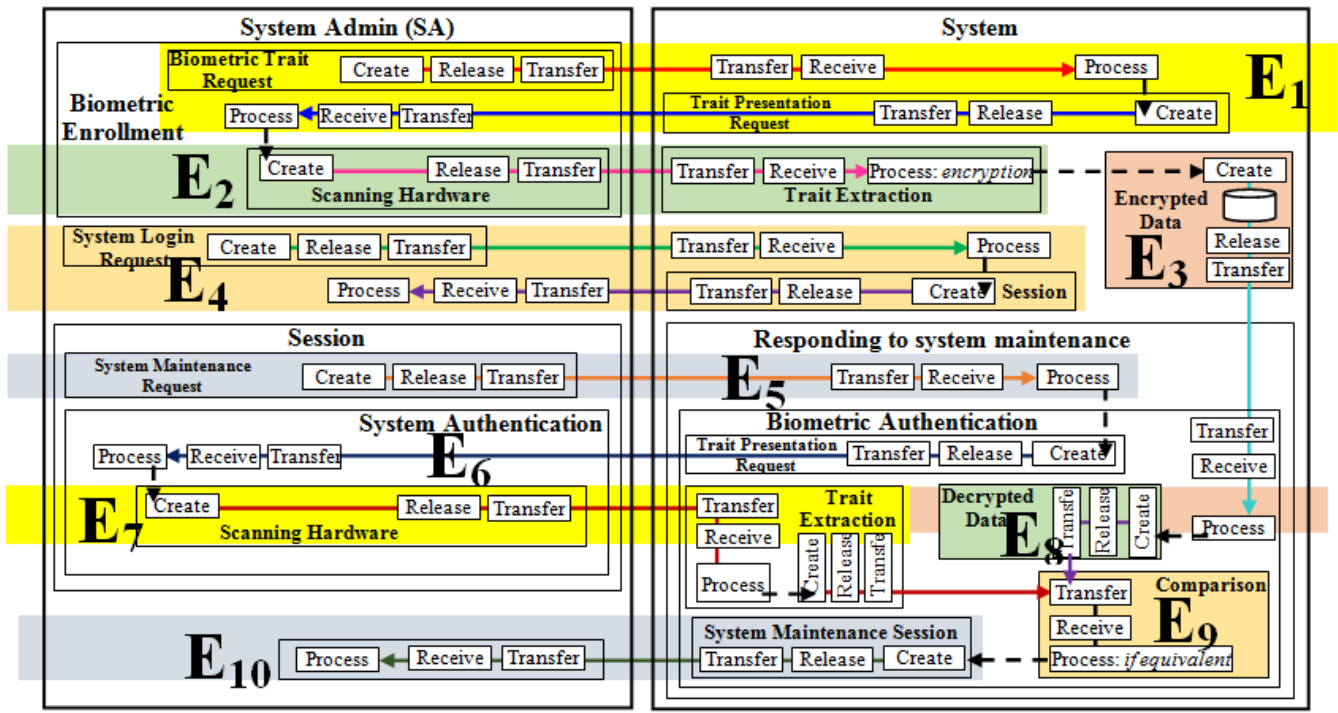

Fig. 18. Meaningful Events During Physical Biometric Authentication.

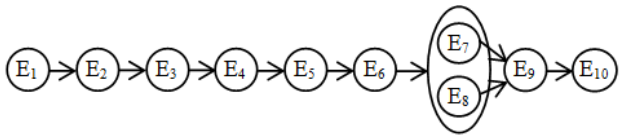

Fig. 19. Control of the Physical Biometric Event Sequence.

Fig. 19 shows the chronology of events modeling the behavior of the biometric authentication system.

### C. Multifactor Authentication

The diagrams for multifactor authentication are not shown for space considerations. A typical multifactor system carries out authentication in two or more stages—the login stage and other verification stage(s) involving other types of authentication.

### D. Multifactor Authentication

The diagrams for multifactor authentication are not shown for space considerations. A typical multifactor system carries out authentication in two or more stages—the login stage and other verification stage(s) involving other types of authentication.

This paper assumes the common choice of randomly generated one-time passwords (OTPs) with two-factor authentication. The TM model comprises two main machines: the SA and the system (highlighted in yellow).

- To initiate an interaction with the system, the SA logs into his or her account. With the correct credentials, the system creates a session.

- The SA issues a request to maintain the system (e.g., software update).

- The system starts the authentication process [36].

    o The system identifies the SA's registered phone number and uses it to generate an OTP.

    o This password is embedded in an SMS and transferred to the system's phone.

    o The system sends the SA an SMS containing the OTP.

    o The SA then inputs the requested OTP in the displayed form.

    o The system extracts the entered OTP for comparison.

    o The OTP entered is compared to the one initially sent. If they are the same, a system maintenance session is opened to the SA.

### VI. Conclusion

In this paper, we presented the thesis that five generic processes—creating, releasing, transferring, receiving, and processing—have the expressive power to model key public infrastructure, biometric, and multifactor authentications. Expressiveness refers to things said in a description in a language [2]. The interesting aspect of the TM is the question of whether TM's five generic processes express all things required in conceptual modeling in software engineering. Indicators including modeling authentication in this paper point to the viability of this hypothesis. Further research should pursue this line of thinking.

### References

[1] G. Kotonya and I. Sommerville, Requirements Engineering: Processes and Techniques. Hoboken: John Wiley & Sons, 1998.

[2] S. Patig, "Measuring Expressiveness in Conceptual Modeling," in Advanced Information Systems Engineering, A. Persson and J. Stirna, Eds. Berlin: Springer, 2004, pp. 127–141 [CAiSE 2004, Lecture Notes in Computer Science, vol. 3084].

[3] R. Y. Lee, "Chapter 4: Modeling with UML," in Object-Oriented Software Engineering with UML: A Hands-On Approach. Hauppauge, NY: Nova Science Publishers, Inc., January 2019.

[4] O. Altuhhova, R. Matulevičius, and N. Ahmed, "Towards definition of secure business process," in Lecture Notes in Business Information Research. Berlin: Springer, 2012, pp. 1–15 [CAiSE 2012 Workshop on Information Systems Security Engineering, 2012].

[5] P. Bresciani, P. Giorgini, F. Giunchiglia, J. Mylopoulos, and A. Perini, "TROPOS: An agent-oriented software development methodology," J. Auton. Agents Multi-Agent Syst, vol. 8, no. 3, pp. 203–236, May 2004.

[6] I. Soomro and N. Ahmed, "Towards security risk-oriented misuse cases," in Business Process Management Workshops. Berlin: Springer, vol. 132, 2013, pp. 689–700.

[7] G. Sindre, Mal-Activity Diagrams for Capturing Attacks on Business Processes. In: Sawyer P., Paech B., Heymans P. (eds) Requirements Engineering: Foundation for Software Quality. REFSQ 2007. Lecture Notes in Computer Science, vol 4542. Springer, Berlin, Heidelberg, pp. pp 355-366, 2007.

[8] Object Management Group, OMG Unified Modeling Language Superstructure. Version 2.2, http://www.omg.org.

[9] J. Evermann, "Thinking ontologically: Conceptual versus design models in UML," in Ontologies and Business Analysis, M. Rosemann and P. Green, Eds., Location: Idea Group Publishing, 2005.

[10] M. Brambilla and P. Fraternali, "Chapter 11: Tools for model-driven development of interactive applications," In View on ScienceDirect Interaction Flow Modeling Language, M. Brambilla, P. Fraternali, and M. Kaufmann (eds.), Elsevier Science, pp. 335-358, 2015.

[11] S. Al-Fedaghi and M. Bayoumi, "Computer attacks as machines of things that flow," 2018 International Conference on Security and Management, Las Vegas, NV, July 30–August 2, 2018.

[12] H. Morris, To Be, To Have, To Know: Smart Ledgers & Identity Authentication, Z/Yen Group, February 2019. https://www.zyen.com/media/documents/To_Be_To_Have_To_Know_Smart_Ledgers__Identity_Authentication.pdf

[13] The Unified Modeling Language, Single Sign-On for Google Apps UML Activity Diagram Example, accessed 3/8/2019. https://www.uml-diagrams.org/google-sign-on-uml-activity-diagram-example.html

[14] H. Storrle and J. H. Hausmann, "Towards a formal semantics of UML 2.0 activities," Software Engineering 2005, vol. P-64 of Lecture Notes on Informatics, Bonn, Germany, pp. 117-128, 2005.

[15] V. Plavsic and E. Secerov, "Modeling of login procedure for wireless application with interaction overview diagrams," Comput. Sci. Inf. Syst. vol. 5, no. 1, pp. 87–108, June 2008.

[16] S. Al-Fedaghi and G. Aldamkhi, "Conceptual modeling of an IP phone communication system: A case study," 18th Annual Wireless Telecommunications Symposium, New York, NY, April 9–12, 2019.

[17] S. Al-Fedaghi and O. Alsumait, "Toward a conceptual foundation for physical security: Case study of an IT department," Int. J. Saf. Secur. Eng., vol. 9, no. 2, pp. 137–156, 2019.

[18] S. Al-Fedaghi and Y. Atiyah, "Modeling with thinging for intelligent monitoring system," IEEE 89th Vehicular Technology Conference: VTC2019-Spring Kuala Lumpur, Malaysia, April 28–May 1, 2019.

[19] S. Al-Fedaghi and Al-Fadhli, "Modeling an unmanned aerial vehicle as a thinging machine," 5th International Conference on Control, Automation and Robotics, Beijing, China, April 19–22, 2019.

[20] S. Al-Fedaghi and E. Haidar, "Programming is diagramming is programming," 3rd International Conference on Computer, Software and Modeling, Barcelona, Spain, July 14–16, 2019.

[21] M. Heidegger, "The thing," in Poetry, Language, Thought, A. Hofstadter, Trans. New York: Harper & Row, 1975, pp. 161–184.

[22] K. Riemer, R. B. Johnston, D. Hovorka, and M. Indulska, "Challenging the philosophical foundations of modeling organizational reality: The case of process modeling," International Conf. on Information Systems, Milan, Italy, 2013. http://aisel.aisnet.org/icis2013/proceedings/BreakthroughIdeas/4/.

[23] S. Al-Fedaghi, "Five generic processes for behaviour description in software engineering," Int. J. Comp. Sci. Inf. Secur., vol. 17, no. 7, July 2019.

[24] S. Al-Fedaghi, "Toward maximum grip process modeling in software engineering," Int. J. Comput. Sci. Inf. Secur., vol. 17, no. 6, June 2019.

[25] L. W. Howe, "Heidegger's discussion of 'the Thing': A theme for deep ecology," Between Species, vol. 9, no. 2, art. 11, 1993. doi:10.15368/bts.1993v9n2.9.

[26] L. R. Bryant, "Towards a machine-oriented aesthetics: On the power of art," paper presented at The Matter of Contradiction Conference, Limousin, France, 2012.

[27] P. Byrd, Generic Meaning, accessed 5/8/2019. http://www2.gsu.edu/~eslhpb/grammar/lecture_5/generic.html

[28] G. Guizzardi and G. Wagner, "Tutorial: Conceptual simulation modeling with onto-UML," Proceedings of the 2012 Winter Simulation Conference, Berlin, Germany, December 9–12, 2012.

[29] N. Nostro, A. Ceccarelli, A. Bondavalli, and F. Brancati, "Insider threat assessment: A model-based methodology," Op. Syst. Rev., vol. 48, no. 2, pp. 3–12, December 2014.

[30] S. M. Dejamfar and S. Najafzadeh, "Authentication techniques in cloud computing: A review," Int. J. Adv. Res. Comput. Sci. Softw. Eng., vol. 7, no. 1, pp. 95–99, January 2017.

[31] A. Banerjee and M. Hasan, Token-Based Authentication Techniques on Open Source Cloud Platforms, Systems and Telematics, Vol. 16, No. 47, pp. 9-29, October-December, 2018.

[32] M. Qasaimeh, R. Turab, R. S. Al-Qassas, Authentication techniques in smart grid: a systematic review, TELKOMNIKA, Vol.17, No.3, pp.1584-1594, June 2019.

[33] A. Agarkar and H. Agrawal, A review and vision on authentication and privacy preservation schemes in smart grid network, Security and Privacy, Vol. 2, No. 2, pp. 1-18,March/April 2019.

[34] M. Furuhed (2018). Public key infrastructure (PKI) explained in 4 minutes, Nexusgroup.com, accessed 5/8/2019. https://www.nexusgroup.com/blog/crash-course-pki.

[35] W. Yang, S. Wang, J. Hu, G. Zheng, and C. Valli, "Security and accuracy of fingerprint-based biometrics: A review," Symmetry, vol. 11, no. 2, art. 141, January 2019. https://www.mdpi.com/2073-8994/11/2/141.

[36] K. Garska (2018). Two-Factor Authentication (2FA) Explained: Email and SMS OTPs, Identity Automation Site, September 27, 2018. https://blog.identityautomation.com/two-factor-authentication-2fa-explained-email-and-sms-otps.